\documentstyle[twocolumn,prb,aps]{revtex}
\begin{document}
\twocolumn[
\title{Renormalization group study of interacting electrons}
\author{Gennady Y. Chitov and David S\'en\'echal}
\address{D\'epartement de Physique and Centre de Recherche en Physique du
Solide,}
\address{Universit\'e de Sherbrooke, Sherbrooke, Qu\'ebec, Canada J1K 2R1.}
\date{May 1995}
\maketitle
\maketitle

\widetext\leftskip=1cm\rightskip=1cm\small
The renormalization-group (RG) approach proposed earlier by Shankar for
interacting spinless fermions at $T=0$ is extended to the case of non-zero
temperature and spin. We study a model with $SU(N)$-invariant short-range
effective interaction and rotationally invariant Fermi surface in  two and
three
dimensions. We show  that the Landau interaction function of the Fermi liquid,
constructed from the  bare parameters of the low-energy effective action, is RG
invariant. On the other hand, the physical forward scattering vertex is found
as
a stable fixed point of the RG flow. We demonstrate that in $d=2$  and 3, the
RG
approach to this model is equivalent to Landau's mean-field treatment of the
Fermi liquid.  We discuss subtleties associated with the symmetry properties of
the scattering amplitude, the Landau function and the low-energy effective
action. Applying the RG to response functions, we find the compressibility and
the spin susceptibility as fixed points.
\bigskip

]

\narrowtext
%==============================================================================
\section{Introduction}
The necessity to better understand the Physics of strongly correlated
fermions, like in high-$T_c$ superconductors or in the fractional quantum Hall
effect, inspired a lot of theoretical efforts to explain the observable
deviations from Fermi liquid (FL) behavior, and to clarify the foundations of
the Landau Fermi liquid theory itself. Much work has been done to elucidate
the microscopic basis of the FL theory in the framework of
quantum field theory\cite{Landau59,AGD}. However, the development of
Renormalization Group (RG) methods to the problem of interacting fermions in
dimension $d$ is relatively new.

So far, the most successful applications of RG methods to interacting fermions
have been achieved in the one-dimensional case, where it was known from exact
solutions that non FL phases exist (e.g.~the Luttinger liquid). For a review on
1D systems see Ref.~\onlinecite{Solyom}. Later, Bourbonnais and
Caron\cite{Bourbonnais91} made an extensive RG study of one-dimensional and
quasi-one-dimensional fermion systems at finite temperature, in which a
transition from Luttinger liquid to FL behavior is revealed. The development
of a RG theory for interacting fermions in isotropic systems of dimensions
greater than one is more recent\cite{Benfatto,Shankar91,Polchinski93,Shankar}.
Such a generic theory, when completed, should include the Landau FL theory as
one of the fixed points of the RG equations. As discussed by Shankar in a very
pedagogical paper,\cite{Shankar} the RG treatment of fermions is
a much more complicated issue than the analogous procedure for critical
phenomena, because of two crucial points: Firstly, the existence of a Fermi
surface: the low-energy modes lie in the vicinity of a continuous geometrical
object (the Fermi surface) and not only around isolated points, like the
origin of phase space in the bosonic case. This introduces additional phase
space constraints on the modes to be integrated out, quite a problem for an
arbitrary Fermi surface. Moreover, the Fermi surface itself is a relevant
parameter of theory, and its shape should renormalize under the RG procedure,
except in the rotationally symmetric case. Secondly, instead of studying the
flow of one or a few coupling constants like in critical phenomena, one has to
set up RG flow equations for coupling {\it functions}, defined on the Fermi
surface. Notice that the purely 1D interacting fermion system can be
considered as a degenerate special case where the Fermi surface is reduced to
a set of two points, with a finite number of coupling constants. It is
therefore closer to the usual applications of the renormalization group
(however, in the quasi-one-dimensional approach of
Ref.~\onlinecite{Bourbonnais91}, one already witnesses the appearance of new
coupling constants related to interchain hopping and one may follow the
change in shape of the (open) Fermi surface under RG flow).

In this paper we extend Shankar's approach\cite{Shankar}, developed for
spinless fermions at zero temperature. The main goal of our analysis is to
incorporate spin and a finite temperature. For the sake of generality, we
study a system of $N$-component fermions with an $SU(N)$-invariant
short-range interaction. We perform the analysis for a circular Fermi surface
in dimension two, spherical in dimension three.

Shankar's analysis of coupling functions for the quartic interaction
shows that only two such functions survive under the RG
flow: the function $\hat\Phi$, which couples two incoming and two
outgoing particles with the same pairs of momenta lying on the Fermi
surface (Landau channel), and the function $\hat V$, which
couples incoming and outgoing
particles with opposite momenta lying on the Fermi surface
(Bardeen-Cooper-Schrieffer (BCS) channel).

We find that the coupling function $\hat V$ is marginally irrelevant if its
bare value satisfies Landau's conditions for the stability of the Fermi liquid
against Cooper pairing at arbitrary momentum; otherwise a BCS instability
occurs at finite temperature. In the Landau channel the distinction between the
Landau interaction function $\hat\Phi$ and the physical forward scattering
vertex $\hat\Gamma$ appears naturally in the finite temperature RG formalism.
The function $\hat\Phi$ is RG invariant at non-zero temperatures, while the
vertex $\hat\Gamma$ is attracted towards a non trivial stable fixed point. In
order for the Fermi liquid to be stable up to zero temperature, we find that
the components of the Landau function must obey conditions which coincide with
Pomeranchuk's stability conditions.\cite{Lifshitz80}

The paper is organized as follows: In Section II we describe the low-energy
effective action which is the starting point of our RG analysis.
In Section III we study the two-dimensional case. This section contains a
description of the key points of our analysis and most of the technical
details. There, we obtain the RG equations for the vertices and their fixed
points. In Section IV we apply the same analysis to $d=3$. In Section V we
calculate the compressibility and the spin susceptibility in this RG
framework.

%==============================================================================
\section{The low-energy effective action}
We treat the problem of interacting fermions at finite temperature in the
standard path integral formalism\cite{Negele88} using Grassmann variables
for Fermi fields. The partition function is given by the path integral
\begin{equation}
\label{Z}
Z =\int {\cal D}\bar\psi{\cal D}\psi~e^{S_0 + S_{int}}
\end{equation}
wherein the free part of the action is
\begin{equation}
\label{S0}
S_0 =\int_{({\bf 1})}\bar\psi_{\alpha}({\bf 1})\left[
i\omega_1+\mu-\epsilon({\bf K}_1)\right]\psi_{\alpha}({\bf 1})
\end{equation}
We introduced the following notation:
\begin{mathletters}
\label{notation}
\begin{eqnarray}
\int_{({\bf i})}~~&\equiv&~~{1\over\beta}\int{d{\bf K}_i\over(2\pi)^d}
\sum_{\omega_i}\\
({\bf i}) ~~&\equiv&~~ ({\bf K}_i,\omega_i)
\end{eqnarray}
\end{mathletters}
where $\beta$ is the inverse temperature, $\mu$ the chemical potential,
$\omega_i$ the fermion Matsubara frequencies and $\psi_{\alpha}({\bf i})$ an
$N$-component Grassmann field with a `flavor' index $\alpha$.
Summation over repeated indices is implicit throughout this paper.
We set $k_B=1$ and $\hbar=1$.
The physically interesting case is of course $N=2$, but the generalization
to $N\ne2$ is simple, and it incorporates automatically the simpler case
of spinless fermions ($N=1$).

The general $SU(N)$-invariant quartic interaction may be written as follows:
\begin{eqnarray}
\label{Sint}
S_{int} &=& -\frac14\int_{({\bf 1},{\bf 2},{\bf 3},{\bf 4})}
\bar\psi_{\alpha}({\bf 1})\bar\psi_{\beta}({\bf 2})
\psi_{\gamma}({\bf 3})\psi_{\delta}({\bf 4}) \nonumber\\
&& \times U^{\alpha\beta}_{\gamma\delta}({\bf 1},{\bf 2};{\bf 3},{\bf 4})
\delta^{(d+1)}({\bf 1}+{\bf 2}-{\bf 3}-{\bf 4})
\end{eqnarray}
Here the conservation of energy and momentum is enforced by the symbolic delta
function
\begin{eqnarray}
\label{Delta}
&&\delta^{(d+1)}({\bf 1}+{\bf 2}-{\bf 3}-{\bf 4})\equiv\beta(2\pi)^d
\times\nonumber\\
&&\delta({\bf K}_1+{\bf K}_2-{\bf K}_3-{\bf K}_4)
\Delta(\omega_1+\omega_2-\omega_3-\omega_4)
\end{eqnarray}
where the discrete delta function $\Delta$ is equal to 1 if its argument is
zero, and equal to zero otherwise.

It can be shown in group theory (see, for instance, Ref.~\onlinecite{Cheng84})
that a representation of the potential as
\begin{equation}
\label{pot1}
U^{\alpha\beta}_{\gamma\delta}
= U_1\delta_{\alpha\delta}\delta_{\beta\gamma}
+ U_2\sum_{a=1}^{N^2-1}\lambda^a_{\alpha\delta}\lambda^a_{\beta\gamma}
\end{equation}
supplies us with the most general $SU(N)$-invariant form with two independent
scalar functions $U_1$ and $U_2$. The $N^2-1$ Hermitian traceless
matrices $\hat\lambda^a$ are the generators of $SU(N)$. We need not
write down explicit expressions of the commutations or other relations for
those matrices. The only identity used in the following is
\begin{equation}
\label{gen}
\sum_{a=1}^{N^2-1}\lambda^a_{\alpha\beta}\lambda^a_{\gamma\delta} =
2\left(\delta_{\alpha\delta}\delta_{\beta\gamma} - {1\over N}
\delta_{\alpha\beta}\delta_{\gamma\delta}\right)
\end{equation}
In the $SU(2)$ case, the three generators $\lambda^a$ are the usual Pauli
matrices and the identity (\ref{gen}) reduces to a well-known relation
involving these matrices. Using the relation (\ref{gen}), one readily
checks that the form (\ref{pot1}) of the interaction is indeed $SU(N)$
invariant. The fact that Eq.~(\ref{pot1}) is the most general
$SU(N)$ invariant may be verified by counting the number of
singlets in the tensor product
$\bar N\otimes\bar N\otimes N\otimes N$, wherein $N$ stands for the
fundamental representation of $SU(N)$ (acting on the $N$-component field
$\psi$) and $\bar N$ for its conjugate. This number is indeed two, meaning
that only two $SU(N)$-invariant scalars may be constructed in this way.

One should bear in mind the difference between the
$SU(N)$-invariant interaction considered here and a rotation invariant
interaction for particles of spin $s$: In the absence of symmetry-breaking
interactions (e.g.~spin-orbit, dipole-dipole, or external
fields) there will be rotational invariance in spin space, but the
corresponding symmetry operations are still obtained from the three
generators of $SU(2)$, although in a
$(2s+1)$-dimensional representation. The $SU(N)$-invariance imposed here is
more stringent. The particular form of Eqs.~(\ref{pot1},\ref{gen}), and
consequently of Eq.~(\ref{potS}), for any $N$, is the artifact of such a
symmetry enlargement. A generic, $SU(2)$-invariant interaction with spin-$s$
fermions would contain a greater variety of terms. Accordingly, the different
components of $\psi$ are called `flavors' if
$N=2s+1>2$, in order to avoid misunderstandings.

The flavor dependence of the potential (\ref{pot1}) may be factorized and
expressed via two independent flavor operators. It is convenient to introduce
two operators $\hat I$ and $\hat T$, respectively symmetric and
antisymmetric, as follows:
\begin{mathletters}
\label{base}
\begin{eqnarray}
I^{\alpha\beta}_{\gamma\delta}~&\equiv&~
\delta_{\alpha\delta}\delta_{\beta\gamma} +
\delta_{\alpha\gamma}\delta_{\beta\delta}\\
T^{\alpha\beta}_{\gamma\delta}~&\equiv&~
\delta_{\alpha\delta}\delta_{\beta\gamma} -
\delta_{\alpha\gamma}\delta_{\beta\delta}
\end{eqnarray}
\end{mathletters}
These operators satisfy the properties
\begin{mathletters}
\label{sym}
\begin{eqnarray}
I^{\alpha\beta}_{\gamma\delta} &=&
\phantom{-}I^{\beta\alpha}_{\gamma\delta} =
\phantom{-}I^{\alpha\beta}_{\delta\gamma}\\
T^{\alpha\beta}_{\gamma\delta} &=&
-T^{\beta\alpha}_{\gamma\delta} = -T^{\alpha\beta}_{\delta\gamma}
\end{eqnarray}
\end{mathletters}
and the convolution relations
\begin{mathletters}
\label{convol}
\begin{eqnarray}
I^{\alpha\beta}_{\mu\nu}I^{\nu\mu}_{\gamma\delta} &=&
2I^{\alpha\beta}_{\gamma\delta}\\
T^{\alpha\beta}_{\mu\nu}T^{\nu\mu}_{\gamma\delta} &=&
2T^{\alpha\beta}_{\gamma\delta}\\
T^{\alpha\beta}_{\mu\nu}I^{\nu\mu}_{\gamma\delta} &=& 0\\
I^{\alpha\mu}_{\gamma\nu}I^{\nu\beta}_{\mu\delta} &=&
\phantom{-}{N+3\over2}I^{\alpha\beta}_{\gamma\delta}
-{N+1\over2}T^{\alpha\beta}_{\gamma\delta}\\
I^{\alpha\mu}_{\gamma\nu}T^{\nu\beta}_{\mu\delta} &=&
-{N-1\over2}I^{\alpha\beta}_{\gamma\delta}
+{N+1\over2}T^{\alpha\beta}_{\gamma\delta}\\
T^{\alpha\mu}_{\gamma\nu}T^{\nu\beta}_{\mu\delta} &=&
\phantom{-}{N-1\over2}I^{\alpha\beta}_{\gamma\delta}
-{N-3\over2}T^{\alpha\beta}_{\gamma\delta}
\end{eqnarray}
\end{mathletters}
Instead of Eq.~(\ref{pot1}), we may decompose the potential as follows:
\begin{equation}
\label{potS}
U^{\alpha\beta}_{\gamma\delta} = U^A I^{\alpha\beta}_{\gamma\delta}
+ U^S T^{\alpha\beta}_{\gamma\delta}
\end{equation}
where the functions $U^S$ and $U^A$ have the symmetry properties
\begin{mathletters}
\label{USYM}
\begin{eqnarray}
U^A({\bf 1},{\bf 2};{\bf 3},{\bf 4}) &=&
-U^A({\bf 2},{\bf 1};{\bf 3},{\bf 4}) =
-U^A({\bf 1},{\bf 2};{\bf 4},{\bf 3})\\
U^S({\bf 1},{\bf 2};{\bf 3},{\bf 4}) &=&
\phantom{-} U^S({\bf 2},{\bf 1};{\bf 3},{\bf 4}) =
\phantom{-} U^S({\bf 1},{\bf 2};{\bf 4},{\bf 3})
\end{eqnarray}
\end{mathletters}
The general form (\ref{Sint}) of the interaction allows us to easily recover
various special cases. Spinless fermions correspond to $N=1$:
matrices have only one component and $\hat  I\equiv 2$,
$\hat T\equiv 0$. Thus there is only one interaction function. For instance,
in the spinless Hubbard model with nearest-neighbor interaction
constant $U^{nn}_0$, the function $U^A$ is expressed in terms of $U^{nn}_0$ and
some combination of trigonometric functions, depending on the spatial
dimension\cite{Shankar}.

In the electron Hubbard model with on-site interaction constant $U^{os}_0$, the
functions are $U^A =0$ and $U^S\propto U^{os}_0$. Switching on a constant
interaction $U^{nn}_0$ between nearest neighbors, we come up with two
independent functions $U^S$ and $U^A$ in the Hamiltonian.

The expressions (\ref{S0}) and (\ref{Sint}) for the action are adequate
for a microscopic, `exact' formulation of the problem. The functions
$U^A$ and $U^S$ may incorporate the microscopic interaction of our
choice: Coulomb, on-site repulsion, and so on. The integration in
Eq.~(\ref{notation}) is then carried over all available phase space
(the Brillouin zone), with the constraint of momentum conservation.
In principle, working with the microscopic Hamiltonian allows a description
of physical processes at all energy scales, up to atomic energies.
However, such an exact solution is impossible in practice. We will consider
the action defined in Eqs.(\ref{S0},\ref{Sint}) as a {\it low-energy
effective action}\cite{Polchinski93,Shankar,Weinberg}, describing correctly
only
those physical processes occurring at an energy scale $\Lambda_0$ much smaller
than the Fermi energy:
\begin{equation}
\label{LKF}
\Lambda_0\ll K_F
\end{equation}
In principle, such a low-energy effective action could be
obtained from the microscopic action by integrating out (in the functional
sense) the degrees of freedom associated with the momenta lying out of a
band of width $2\Lambda_0$ around the Fermi surface.

In general, the low-energy effective action $S_0+S_{int}$ is written as an
expansion in a set of Grassmann fields, with the requirements that the
symmetries of the microscopic action be satisfied (for details, see Refs.
\onlinecite{Polchinski93,Shankar}). We assume that the effective action has
the form given in Eqs~(\ref{S0}) and (\ref{Sint}), except that the integration
in Eq.~(\ref{notation}) is restricted to the vicinity of the Fermi surface.
The parameters of this action, like one-particle energy $\epsilon(\bf K)$ and
the coupling functions $U^A$ and $U^S$, do not coincide with those of the
microscopic action in general. We ignore the possibility of interaction terms
involving derivatives, or more powers of $\psi$, because such terms are
irrelevant at tree-level (see below).

In this work we will restrict ourselves to the study of low-density
electron systems, so we assume that the Fermi surfaces have rotational
symmetry (i.e.~circular or spherical). The low-lying one-particle
excitations can then be linearized near the Fermi energy in a simple fashion:
\begin{equation}
\label{linear}
\epsilon({\bf K})-\mu\approx v_F(K-K_F)\equiv v_Fk
\end{equation}
wherein the terms of order $k^2$ in this expansion are irrelevant.
The momentum $k$ and the temperature $T$ are restricted by the inequalities
$|k|\leq\Lambda_0$ and $T\ll v_F\Lambda_0$ along with the condition
(\ref{LKF}).
The bare one-particle Green's function for the free part of action $S_0$ is
\begin{eqnarray}
\label{G0def}
\langle\psi_{\alpha}({\bf 1})\bar\psi_{\beta}({\bf 2})\rangle_0 &=&
G_0({\bf K}_1,\omega_1)
\delta^{(d+1)}({\bf 1}-{\bf 2})\delta_{\alpha\beta}\\
\label{G0}
G_0^{-1}({\bf K}_1,\omega_1) &=& i\omega_1+\mu-\epsilon({\bf K}_1)
\simeq i\omega_1-v_Fk_1
\end{eqnarray}
In the $d$-dimensional integration measure only the relevant term is kept:
\begin{eqnarray}
\label{measure}
\int d{\bf K}&~=~&\int_{-\Lambda_0}^{\Lambda_0}
\int_{\Omega_d}{(K_F+k)}^{d-1}dkd\Omega_d \nonumber\\ &~\simeq~&
K_F^{d-1}\int_{-\Lambda_0}^{\Lambda_0}\int_{\Omega_d} dkd\Omega_d
\end{eqnarray}
The Matsubara frequencies are allowed to run over all available values. We
presume that the density of particles in the system is kept fixed.

The low-energy effective action (\ref{S0},\ref{Sint}) lends itself to a
simple tree-level scaling analysis (dimensional analysis). It is invariant
under the rescaling $\omega_n^{\prime}= s\omega_n$, $k^{\prime} =sk$ and
$\psi^{\prime}(k^{\prime},\omega^{\prime}_n)= s^{-3/2}\psi(k,\omega_n)$.
Higher orders in the expansion (\ref{linear}) or in the measure
(\ref{measure}) are then irrelevant (in the RG sense) at tree-level, and so
are more complicated interaction monomials. This justifies the simple form
(\ref{S0},\ref{Sint}).

According to the RG analysis of Shankar\cite{Shankar}, performed at zero
temperature, this effective action describes the Fermi liquid phase of
repelling spinless fermions and the BCS superconducting phase of attracting
spinless fermions. We will study here the RG
fixed points for interacting $N$-flavor fermions expanding this technique
for the finite temperature case.

%==============================================================================
\section{The RG equations in two dimensions }
%------------------------------------------------------------------------------
\subsection{The coupling functions}
In this section we will restrict
the analysis to the two-dimensional case, with a circular Fermi surface.
As follows from tree-level scaling, the frequency
dependence of the coupling function $\hat U$ is irrelevant and is
discarded from now on.

Each vector ${\bf K}_i$ in the effective action lies in a narrow shell of
thickness $2\Lambda_0$ around the Fermi surface. We may write it as
${\bf K}_i={\bf K}_F^i+{\bf k}_i$ where the vector ${\bf k}_i$ is orthogonal to
the Fermi surface and small, in the sense that
$|{\bf k}_i|\leq\Lambda_0\ll K_F$. The coupling function is then written as
\begin{equation}
\label{Uq}
U^{A,S}= U^{A,S}({\bf K}_F^1,{\bf K}_F^2;{\bf K}_F^3,{\bf K}_F^4|
{\bf k}_1,{\bf k}_2;{\bf k}_3,{\bf k}_4)
\end{equation}
If we restrict ourselves to the case of non-singular
interactions,\cite{Houghton94} the
$\bf k$-dependence of the coupling function turns out to be
irrelevant. Moreover, the dependence of the coupling function on the
Fermi surface momenta ${\bf K}_i$ is constrained by crystal momentum
conservation: ${\bf K}_F^1+{\bf K}_F^2={\bf K}_F^3+{\bf K}_F^4$ (we assumed
that Umklapp processes are inoperative, which is justified for a circular
Fermi surface at low-filling).

A geometrical analysis (see Ref.\onlinecite{Shankar}) leads to the following
conclusion: In the case of a circular Fermi surface, the constraint that the
momenta ${\bf K}_F^i$ lie on the Fermi surface and the momentum conservation
law allow the existence of two pairs of independent coupling functions, as
described below.

{\it Case 1.~}
If ${\bf K}_F^1\ne -{\bf K}_F^2$, we define a dimensionless coupling
function $\hat\Phi$ of the momenta and spins:
\begin{equation}
\label{Fdef}
\hat\Phi({\bf K}_F^1,{\bf K}_F^2) \equiv\frac12\nu_F
U^{\alpha\beta}_{\gamma\delta} ({\bf K}_F^1,{\bf K}_F^2;
{\bf K}_F^1,{\bf K}_F^2)
\end{equation}
where $\nu_F = 2S_dK_F^{d-1}/{(2\pi)}^d v_F$ is the density of
electron states at the Fermi level  and $S_d$ is the area of the
$d$-dimensional unit sphere. The hat ($\hat{~}$) means that $\hat\Phi$ is an
operator in flavor space. Each vector ${\bf K}_F^a$ may be specified by a
plane polar angle $\theta_a$; because of rotation invariance, the function
$\hat\Phi$ may only depend on the relative angle $\theta_1-\theta_2$
between ${\bf K}_F^1$ and ${\bf K}_F^2$:
\begin{equation}
\hat\Phi(\theta_1-\theta_2)~=~\Phi^A(\theta_1-\theta_2)
I^{\alpha\beta}_{\gamma\delta} ~+~\Phi^S(\theta_1-\theta_2)
T^{\alpha\beta}_{\gamma\delta}
\end{equation}
It should be pointed out that the function $\Phi^A(\theta)$
is not an antisymmetric function of its argument. It follows from the symmetry
properties of the coupling function (Eqs.~(\ref {sym},\ref {potS},\ref{USYM}))
that
\begin{equation}
\label{FasSym}
\Phi^{A,S}(\theta_1-\theta_2)=\Phi^{A,S}(\theta_2-\theta_1)
\end{equation}
The only remnant of the antisymmetry of $U^A$ is the  condition $\Phi^A(0)=0$.
It is easy to check that an exchange of incoming or outgoing momenta does not
require any new independent function.

{\it Case 2.~}
The two outgoing momenta are opposite:
${\bf K}_F^2 = -{\bf K}_F^1$ (likewise for the two incoming momenta).
We introduce another pair of dimensionless functions:
\begin{mathletters}
\label{Vas}
\begin{eqnarray}
\hat V({\bf K}_F^1,{\bf K}_F^3) &\equiv &\frac12\nu_F
U^{\alpha\beta}_{\gamma\delta}({\bf K}_F^1,-{\bf K}_F^1;{\bf K}_F^3,-{\bf K}_
F^3)\\
&=&V^A(\theta_{13})  I^{\alpha\beta}_{\gamma\delta} +
V^S(\theta_{13}) T^{\alpha\beta}_{\gamma\delta}
\end{eqnarray}
\end{mathletters}
wherein $\theta_{13}\equiv\theta_1-\theta_3$.
{}From the symmetry properties of the coupling function $\hat U$ under
exchange of momenta, we find
\begin{mathletters}
\label{VSym}
\begin{eqnarray}
V^A(\theta\pm\pi)&=&-V^A(\theta)\\
V^S(\theta\pm\pi)&=&\phantom{-}V^S(\theta)
\end{eqnarray}
\end{mathletters}
There is an overlap in the definitions of $\hat\Phi$ and
$\hat V$, since $\hat\Phi(\pm\pi)={\hat V}(0)$
or, more explicitly,
\begin{mathletters}
\label{pi}
\begin{eqnarray}
\Phi^A(\pm\pi)=V^A(0)&=&-V^A(\pm\pi)\\
\Phi^S(\pm\pi)=V^S(0)&=&\phantom{-}V^S(\pm\pi)
\end{eqnarray}
\end{mathletters}
Because of the $2\pi$-periodicity of the functions $\Phi^{A,S}$,
$V^{A,S}$ and of the symmetry  properties (\ref{FasSym}) and (\ref{VSym}),
these functions take all their possible values in the following `minimal'
domains: $\theta\in [0,\pi)$ for $\Phi^{A,S}(\theta)$ and
$\theta\in [0,\pi]$ for $V^{A,S}(\theta)$. The angle $\theta=\pi$ is excluded
from the domain of $\Phi^{A,S}$ in order to avoid any ambiguity following from
(\ref{pi}).

In the rotationally invariant case it is convenient to expand
the coupling functions in Fourier series:
\begin{mathletters}
\label{Fourier}
\begin{eqnarray}
{\bf X}(\theta) ~&=&~
\sum_{l=-\infty}^{\infty}e^{-il\theta}{\bf X}_l\\
{\bf X}_l~&=&~\int_0^{2\pi}{d\theta\over 2\pi}~{\bf X}
(\theta)e^{il\theta}
\end{eqnarray}
\end{mathletters}
where ${\bf X}$ stands for the set of all coupling functions:
${\bf X}=\{\Phi^A,\Phi^S, V^A, V^S\}$.
In terms of Fourier components, the symmetry properties of $\hat\Phi$
become $\Phi^{A,S}_l=\Phi^{A,S}_{-l}$ together with
\begin{equation}
\label{Pauli2}
\sum_{l=-\infty}^{\infty}\Phi^A_l=0
\end{equation}
Those of $\hat V$ become $V^A_l=0$ ($l$ even) and $V^S_l=0$ ($l$ odd).
The symmetry properties of the coupling functions expressed above are exact,
for they are consequences of the Pauli principle.

Let us finally point our that the functions $\Phi^A$ and $V^A$
coincide with those introduced by Shankar\cite{Shankar} in the spinless
case, while $\Phi^S$ and $V^S$ are brought about by the
introduction of spin.

%------------------------------------------------------------------------------
\subsection{RG equations in the Landau channel}
Consider the two-particle Green's function:
\begin{equation}
\label{G2}
{\hat G}_2({\bf 1},{\bf 2};{\bf 3},{\bf 4}) = -
\langle\psi_{\alpha}({\bf 1})\psi_{\beta}({\bf 2})
\bar\psi_{\gamma}({\bf 3})\bar\psi_{\delta}({\bf4})\rangle
\end{equation}
The vertex function
${\hat\Gamma}({\bf 1},{\bf 2};{\bf 3},{\bf 4})$
associated to
${\hat G}_2({\bf 1},{\bf 2};{\bf 3},{\bf 4})$ is
constructed by considering only connected one-particle-irreducible (1PI)
diagrams with amputated external legs. In the first-order
approximation, ${\hat\Gamma}^{(1)}({\bf 1},{\bf 2};{\bf 3},{\bf 4}) =
{\hat U}({\bf 1},{\bf 2};{\bf 3},{\bf 4})$.
% Here again a hat ($\hat{~}$) stands for the flavor indices.

%xxxxxxxxxxxxxxxxxxxxxxxxxxxxxxxxxxxxxxxxxxxxxxxxxxxxxxxxxxxxxxxxxxxxxxxxxx
\begin{figure}[tpb]
\input epsf.tex
\vglue 0.4cm\epsfxsize 8cm\centerline{\epsfbox{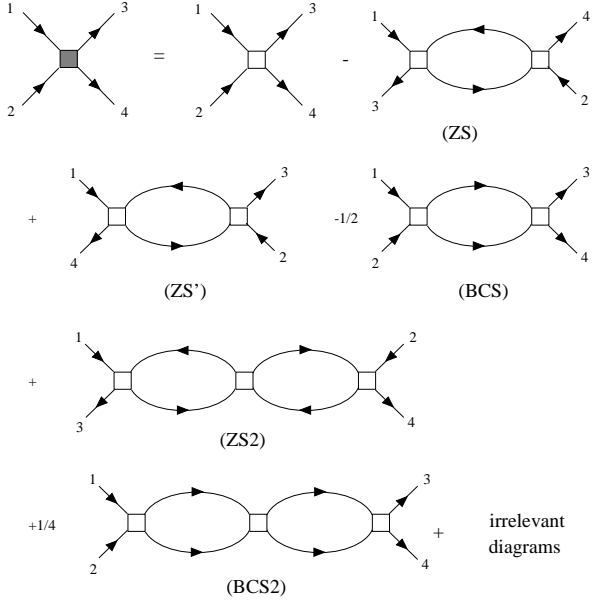}}\vglue 0.4cm
\caption{
Perturbative diagrammatic expansion for the total vertex in terms of the
bare interaction. At the two-loop level, only the diagrams employed in the
Sec.IIID are depicted. }
\end{figure}
%xxxxxxxxxxxxxxxxxxxxxxxxxxxxxxxxxxxxxxxxxxxxxxxxxxxxxxxxxxxxxxxxxxxxxxxxxx
Let us define a flow parameter $t$ such that $\Lambda(t)=\Lambda_0e^{-t}$. We
will follow the flow of the vertex function from $t=0$ to $t=\infty$. We
introduce the dimensionless temperature
$\beta_R(t)\equiv\frac12 v_F\beta\Lambda (t)$. Three Feynman diagrams
contribute to the Gell-Mann--Low (GML) function of the vertex function at the
one-loop level (see Fig.~1): zero sound (ZS), Peierls (ZS$'$),  and Cooper
pairing (BCS). The coupling functions $\hat\Phi$ and ${\hat  V}$ can be
extracted from the vertex ${\hat\Gamma}$ by proper choice of the external
momenta lying on the Fermi surface and by putting the external frequencies
equal to the minimal Matsubara frequency $\omega_{min}\equiv\pi T$.
Summations over flavor indices and Matsubara frequencies can be done easily
using Eq.~(\ref{convol}) and standard techniques\cite{AGD,Negele88} for
working with temperature Green's functions. The relevant contributions
to the GML functions of $\hat\Phi$ and $\hat V$ come only from those  diagrams
in which all the momentum phase space is available for integration.  For a
detailed discussion of this last issue, see Ref.~\onlinecite{Shankar}.

To our knowledge, Shankar was the first to argue\cite{Shankar} that the RG
fixed  point in the Landau channel yields the Landau $\hat f$-function of the
Fermi liquid. At zero temperature, Shankar obtained an RG-invariant function
$\Phi^A$ and identified it with the Landau function. Our finite-temperature
analysis reveals a more subtle situation.

As is known from the Landau FL theory\cite{Landau59}, there are two  limits of
the  four-point vertex function with all four momenta lying on the Fermi
surface, depending on the way the limit of zero momentum- and
energy-transfer is taken. One of these limits, which does not possess the
symmetry (\ref{Pauli2}) of the forward scattering amplitude , yields the Landau
function. The other one, which preserves the antisymmetry of the vertex, gives
the total forward  scattering amplitude.  For details see, for instance,
the especially elucidative paper by Mermin.\cite{Mermin67}

To clarify this point in our approach, let us consider the perturbative
contribution of the ZS diagram to the vertex function ${\hat\Gamma}({\bf
1},{\bf 2};{\bf 3},{\bf 4})$ at small momentum transfer ${\bf q}=
{\bf K}_3-{\bf K}_1$ and small energy transfer
$i\Omega_{13}= i(\omega_1-\omega_3)$.
As discussed after Eq.~(\ref{Uq}), we neglect the difference
${\bf K}^i-{\bf K}^i_F$ in the effective interaction. For
this contribution we have:
\begin{eqnarray}
\label{IZS}
&&{\hat {\cal I}}_{ZS}({\bf K}_F^1,{\bf K}_F^2,{\bf q},\Omega_{13}) =
\int {d{\bf K}\over (2\pi)^d}
U^{\alpha\mu}_{\gamma\nu}({\bf K}_F^1,{\bf K}_F;{\bf K}_F^1,{\bf K}_F)
\nonumber\\ &&\quad\times
U^{\nu\beta}_{\mu\delta}({\bf K}_F,{\bf K}_F^2;{\bf K}_F,{\bf K}_F^2)
L_{ZS}({\bf K},{\bf q},\Omega_{13})
\end{eqnarray}
where
\begin{eqnarray}
\label{LZS}
&&L_{ZS}({\bf K},{\bf q},\Omega_{13}) =
\frac12 {\sinh(\beta v_F{\bf K}\cdot{\bf q}/2K_F)\over  i\Omega_{13} -
v_F{\bf K}\cdot{\bf q}/K_F }\nonumber\\
&&\quad \times \cosh^{-1}\left[\frac12\beta v_F k\right]
\cosh^{-1}\left[{\beta v_F\over 2K_F}
{\bf K}_F\cdot({\bf k}+{\bf q})\right]
\end{eqnarray}
is the result of the summation of the product of two Green's functions over
Matsubara frequencies inside the loop. As follows from Eq.~(\ref{LZS}),
$L_{ZS}$ has two limits: In the `unphysical' limit (${\bf q}\to0$,
followed by $\Omega_{13}=0$) $L_{ZS}$ vanishes. In the `physical' limit
($\Omega_{13}=0$, followed by  ${\bf q}\to0$),
$L_{ZS}=-\frac14\beta\cosh^{-2} (\frac12\beta v_F k)$.

All the phase space is available for integration in the r.h.s.~of
Eq.~(\ref{IZS}),  so the ZS diagram gives a relevant contribution to the
total  vertex at any angle $\theta$ between ${\bf K}_F^1$ and ${\bf K}_F^2$.
A similar analysis for the ZS$'$ diagram shows, that  because of the phase
space restrictions, it is relevant only when  $\theta\to0$, and its
contribution does not depend on the way how zero  energy- and
momentum-transfer limit is taken:
\begin{eqnarray}
\label{IZSpr}
&&\lim_{\theta\to0 }~{\hat {\cal I}}_{ZS'}(0,0) =\int {d{\bf K}\over (2\pi)^d}
U^{\alpha\mu}_{\delta\nu}({\bf K}_F^1,{\bf K}_F;{\bf K}_F^1,{\bf K}_F)
\nonumber\\ &&\quad\times
U^{\nu\beta}_{\mu\gamma}({\bf K}_F,{\bf K}_F^1;{\bf K}_F,{\bf K}_F^1)
\frac14\beta\cosh^{-2} (\frac12\beta v_F k)
\end{eqnarray}
In the following we will call $\hat\Phi(\theta,t)$ the
renormalized dimensionless vertex in the `unphysical' limit, i.e.
\begin{equation}
\label{FiR}
\hat\Phi(\theta,t)\equiv\frac12\nu_F
\left[\lim_{ {\bf q}\to0}
{\hat\Gamma}({\bf 1},{\bf 2};{\bf 3},{\bf 4})\right]_{\Omega_{13}=0}
\end{equation}
This notation is justified since the vertex function coincides with the
effective interaction $\hat\Phi$ in that limit. On the other hand, we
define the renormalized dimensionless vertex in the `physical' limit as
\begin{equation}
\label{GammaR}
{\hat\Gamma}(\theta,t)\equiv\frac12\nu_F
\lim_{ {\bf q}\to0}
\left[{\hat\Gamma}({\bf 1},{\bf 2};{\bf 3},{\bf 4})
\bigg\vert_{\Omega_{13}=0}\right]
\end{equation}
We  will use the components of the vertex $\Gamma^{A,S}$ defined in the same
fashion as in Eq.~(\ref{potS}).

Note that among one-loop
contributions to the  GML function of the vertex $\hat\Gamma$
with arbitrary incoming and outgoing external momenta and frequencies, the
BCS channel possesses the total symmetry of the interaction
(Eqs.~(\ref{USYM})), while the ZS (or ZS$'$) channel separately do not:
only their combined contribution is antisymmetric under
exchange of incoming (or outgoing) particles. Here, the diagram ZS$'$
contributes  to the GML function of $\hat\Phi(\theta,t)$
when $\theta\to0$, and it generates the flow for
$\Phi^A(\theta,t)$ at $\theta=0$, breaking the condition $\Phi^A(0)=0$,
 while this ZS$'$ contribution for the physical vertex cancels
$\partial\Gamma^A(\theta,t)/\partial t$ at $\theta=0$. This makes the
derivatives $\partial\Phi^A(\theta,t)/\partial t$ and
$\partial\Gamma^A(\theta,t)/\partial t$ discontinuous at $\theta =0$.
In reality, this feature is smeared over some
angle $\theta_{sm}\sim\Lambda_0 / K_F$, which is small ($\theta_{sm}\ll 1$)
because of the condition (\ref{LKF}). In order to obtain a smooth
GML functions for $\hat\Phi(\theta,t)$ and $\hat\Gamma(\theta,t)$,
we adopt the following procedure: First, we define $\hat\Phi(\theta,t)$ and
$\hat\Gamma(\theta,t)$
in the patch $\theta\in [\theta_0,\pi)$, where $\theta_{sm}<\theta_0\ll 1$.
Then we derive the RG equations in this domain, and
finally we analytically continue $\hat\Phi(\theta,t)$ and
$\hat\Gamma(\theta,t)$ up to $\theta_0 =0$.

The calculations are straightforward, and we end up with the following RG
equations:
\begin{mathletters}
\label{RGeq}
\begin{eqnarray}
&& {\partial\Phi^{A,S}_l\over\partial t} = 0\\
&& {\partial\Gamma^{A,S}_l\over\partial t}
= {\beta_R\over\cosh^2\beta_R}  P^{A,S}_l\\
&& {\partial\beta_R\over\partial t}=-\beta_R
\end{eqnarray}
\end{mathletters}
Here we introduced the following notation:
\begin{mathletters}
\label{AS}
\begin{eqnarray}
P^A_l &\equiv &~~{N+3\over 2}[\Gamma^A_l]^2-(N-1)\Gamma^A_l\Gamma^S_l+
{N-1\over 2}[\Gamma^S_l]^2\\
P^S_l &\equiv &-{N+1\over 2}[\Gamma^A_l]^2+(N+1)\Gamma^A_l\Gamma^S_l-
{N-3\over 2}[\Gamma^S_l]^2
\end{eqnarray}
\end{mathletters}
As follows from Eqs.~(\ref{RGeq}a), the functions $\Phi^{A,S}(\theta,t)$ are
RG invariant.
The effect of the singular contribution to the RG flow from the ZS$'$ diagram
is not included in Eq.~(\ref{RGeq}a), but is equivalent to relaxing the
symmetry condition $\Phi^A(0)=0$ for the bare coupling function
$\Phi^A(\theta)$. It is  convenient to express the RG invariants $\Phi^{A,S}$
in terms of  `charge' ($F$) and `flavor' ($G$) functions, defined as follows:
\begin{mathletters}
\label{ChFl}
\begin{eqnarray}
F &\equiv &(N-1)\Phi^S-(N+1)\Phi^A\\
G &\equiv &-\Phi^S-\Phi^A
\end{eqnarray}
\end{mathletters}
The physical meaning of these functions becomes clear in the
case of real electron spin ($N=2$), when they coincide with the
components of the Landau $\hat f$-function\cite{Lifshitz80}:
\begin{equation}
\label{FLandau}
f_{12,34}(\theta)= F^{(L)} (\theta)\delta_{13}\delta_{24}+
 G^{(L)} (\theta)\mbox{\boldmath $\sigma$}_{13}\cdot
\mbox{\boldmath $\sigma$}_{24}
\end{equation}

In order to decouple the Eqs.~(\ref{RGeq}b) we introduce new components of the
renormalized vertex ${\hat\Gamma}$:
\begin{mathletters}
\label{ABV}
\begin{eqnarray}
A &\equiv &(N-1)\Gamma^S-(N+1)\Gamma^A\\
B &\equiv &-\Gamma^S-\Gamma^A
\end{eqnarray}
\end{mathletters}
Their RG fixed points in the case $N=2$ coincide with the components of the
total scattering amplitude in the notations of FL theory.\cite{Lifshitz80}
(Notice, that the components $\Gamma^S$ and $-\Gamma^A$ are respectively
called singlet and triplet amplitudes in Ref.~\onlinecite{BaymP}).
Notice also the symmetries $A_l=A_{-l}$ and $B_l=B_{-l}$.
To simplify the system of Eqs.~(\ref{RGeq}b) and (\ref{RGeq}c) we use an
auxiliary RG parameter
\begin{equation}
\label{tau}
\tau\equiv\tanh\beta_R~,~~~\tau\in [0,\tau_0]
\end{equation}
wherein $\tau_0\equiv\tanh\beta_0\leq 1$, and
$\beta_0\equiv\frac12 v_F\beta\Lambda_0$.

Expressed in terms of these new variables, the RG equations become
much simpler:
\begin{mathletters}
\label{RGChFl}
\begin{eqnarray}
{\partial A_l\over\partial\tau} &=&  A_l^2\\
{\partial B_l\over\partial\tau} &=&  B_l^2
\end{eqnarray}
\end{mathletters}
{}From Eqs.~(\ref{RGChFl}) we obtain the following stable fixed points
(i.e.~the
solutions at  $t=\infty$) :
\begin{mathletters}
\label{Fix1}
\begin{eqnarray}
A_l^{\ast} &=& {F_l\over
1+F_l\tau_0 }\\
B_l^{\ast} &=& {G_l\over
1+G_l\tau_0 }
\end{eqnarray}
\end{mathletters}
where we expressed the bare values of the vertex via parameters of the
effective
interaction (the FL parameters), i.e.~$A_l^{(0)}=F_l$, $B_l^{(0)}=G_l$.
(In this work asterisks and zeros respectively denote fixed points and bare
values.) From Eq.~(\ref{Fix1}) it follows that, if the parameters of
interaction satisfy the conditions
\begin{equation}
\label{Stab2}
\{ F_l, G_l\} > -1~,\qquad\forall~ l
\end{equation}
the system will remain stable in the Landau channel at
any temperature.
% Otherwise, the general stability of  the system can be broken below the
% temperature
% \begin{equation}
% \label{Tgap}
% T^{(l)}_g = {v_F\Lambda_0\over\ln[(x_l-1)/(x_l+1)]}
% \end{equation}
% when a pole in the vertex function appears. Here $x_l\equiv min\{
% F_l,G_l\} $, in which only harmonics violating the
% condition (\ref{Stab2}) are included. Notice that if $x_l=-1$, a gap
% opens exactly at zero temperature, while the critical temperature
% $T^{(l)}_g$ increases with decreasing $x_l$ thereafter.
The fixed point (\ref{Fix1})  at $T=0$ ($\tau_0=1$) gives the same solution
for the forward scattering vertex as the  Bethe-Salpeter equation in the
zero-temperature technique.\cite{AGD,Lifshitz80}
Notice that, because of the condition $T\ll v_F\Lambda_0$ imposed on the
low-energy effective action, we can set $\tau_0=1$ for all practical purposes.
Any attempt to extract a critical temperature from Eq.~\ref{Fix1} by looking
at the violation of condition (\ref{Stab2}) for the absence of poles in the
vertex function, would involve exponentially small corrections to
zero temperature, thus exceeding the accuracy of the present approach.

To take into  account the RG flow from the ZS$'$ diagram, which preserves the
symmetry condition $\Gamma^A(\theta=0,t=\infty)=0$, we impose the constraint
\begin{equation}
\label{SUMR}
\sum_{l=-\infty}^{\infty}
\left( (N-1)A_l^{\ast}+ B_l^{\ast}\right) =0
\end{equation}
on the fixed point of the vertex. This specific form (\ref{SUMR}) of the
condition for the forward scattering amplitudes is usually called
{\it the amplitude sum rule} in FL theory.\cite{AGD,Lifshitz80}

%------------------------------------------------------------------------------
\subsection{The RG equations in the BCS channel}
The only relevant contribution to the GML function
for $\hat V(\theta,t)$ comes from the BCS diagram. The fact that the BCS
diagram preserves by itself the symmetry (\ref{VSym}) of the function $\hat V$
makes life simpler. The RG equations are derived easily:
\begin{mathletters}
\label{RGV}
\begin{eqnarray}
{\partial V^A_l\over\partial t} &~=~&\phantom{-}\tanh\beta_R ~(V^A_l)^2
\qquad\hbox{($l$ odd)}\\
{\partial V^S_l\over\partial t} &~=~& -\tanh\beta_R ~(V^S_l)^2
\qquad\hbox{($l$ even)}
\end{eqnarray}
\end{mathletters}
At zero temperature, Eq.~(\ref{RGV}a) coincides with Shankar's
result\cite{Shankar} for spinless fermions.\cite{Note1}

The RG equations~(\ref{RGV}) may be solved easily, and their
fixed points are:
\begin{mathletters}
\label{VFix}
\begin{eqnarray}
V^{A(\ast)}_l &=& {V^{A(0)}_l\over
1-V^{A(0)}_l\int_0^{\infty}\tanh(\beta_0 e^{-t})dt}\\
V^{S(\ast)}_l &=& {V^{S(0)}_l\over
1+V^{S(0)}_l\int_0^{\infty}\tanh(\beta_0 e^{-t})dt}
\end{eqnarray}
\end{mathletters}
The system remains stable in the $\hat V$ channel at any temperature
if, for all harmonics, the following conditions are fulfilled:
\begin{mathletters}
\label{VStab}
\begin{eqnarray}
V^{A(0)}_l &<& 0\\
V^{S(0)}_l &>& 0
\end{eqnarray}
\end{mathletters}
At low temperature (i.e., when the low-energy
action approach is supposed to work well), $\beta_0\gg1$ and the integrals
of Eqs.~(\ref{VFix}) can be evaluated exactly, like in the theory of
superconductivity (cf. section 33.3 of Ref.~\onlinecite{AGD}):
\begin{mathletters}
\label{VFixBCS}
\begin{eqnarray}
V^{A(\ast)}_l &=& {V^{A(0)}_l\over
1-V^{A(0)}_l\ln(2\tilde\gamma v_F\Lambda_0/\pi T)}\\
V^{S(\ast)}_l &=& {V^{S(0)}_l\over
1+V^{S(0)}_l\ln(2\tilde\gamma v_F\Lambda_0/\pi T)}
\end{eqnarray}
\end{mathletters}
wherein $\ln\tilde\gamma\equiv\gamma\approx 0.577$ is Euler's constant.
If the conditions (\ref{VStab}) are violated, the $\hat V$-interaction becomes
marginally relevant, and a pole appears at temperature
\begin{equation}
\label{TBCS}
T^{(l)}_{SC}={2\tilde\gamma\over\pi} v_F\Lambda_0\exp (-{1\over\Xi_l})
\end{equation}
Here $\Xi_l\equiv max\{ V^{A(0)}_l,|V^{S(0)}_l|\}$, in which only the
harmonics violating the condition (\ref{VStab}) are included.
If phonons provide the underlying mechanism of attraction, we may identify the
characteristic energy scale $v_F\Lambda_0$ of the low-energy action with the
Debye energy $\omega_D$. In the present context, the BCS theory of
superconductivity corresponds to the special case of a
contact attractive interaction, for which only the zeroth harmonic
$V^{S(0)}_0\propto U_0<0$ does not vanish.

%-----------------------------------------------------------------------------
\subsection{Beyond the one-loop approximation}
The main goal of this subsection is to argue that there are no higher-loop
corrections neither into one-particle Green's function, nor into the vertex. In
other words, the one-loop RG results are  exact for this model, based on the
low-energy effective action.\cite{Shankar}

The phase space analysis of the diagrams contributing to the self-energy
$\Sigma$ shows, that the only relevant contribution comes from
the tadpole  diagram at the one-loop level, and diagrams obtained from it by
tadpole self-energy insertions at higher loops. This self-energy
contribution is temperature-independent, and it
results in a mere shift of chemical potential of the interacting system
$\mu_{int}$. As far as we retain the constant density of particles in the
system, the self-energy contribution exactly compensates the change of the
chemical potential due to interaction at $T=0$, i.e. $\mu_{int}(T=0)-\Sigma
=\mu=K_F^2/m^{\ast}$, according to the Luttinger theorem.\cite{Lutt60} So,
the renormalization of the one-particle Green's function due to the
self-energy holds its linearized form (\ref{G0}) near the Fermi surface
unchanged. It supposes also the temperatures under consideration being
small enough to neglect the temperature corrections to the chemical
potential of the interacting system.

Thus, integration over the rest of the degrees of freedom inside  the narrow
shell around the Fermi surface neither generates an additional wavefunction
renormalization (we have chosen the initial scale of the Grassmann
fields in the bare effective action such that $Z=1$), nor an additional
renormalization of the effective mass of  quasiparticles
(i.e.~the phenomenological parameter of the effective action
$v_F=K_F/m^{\ast}$ does not change).

To show that the RG results obtained above are exact for this model, we will
demonstrate the cancellation of the next terms in the GML functions at the
two-loop level. The generalization of this proof to higher loops is
straightforward. In this subsection we use the standard field-theory
renormalization technique\cite{BLZ,Amit} with momentum integrations
taken over the interval $k\in [-\Lambda,\Lambda]$, instead of the
Wilson-Kadanoff (WK) scheme. Since we work at
finite temperature, there is no need to introduce an infrared cutoff, because
all integrals are regular near the point $k=0$.
The pictorial form of the equation for the renormalized vertex in terms of the
1PI-diagrams is presented in Fig.~1. The light squares stand for the
bare  vertices, while the dark square denotes the renormalized vertex. The
equation for the renormalized vertex may be written in the symbolic
short-hand form as:
\begin{equation}
\label{UR}
u_R=u_0-u_0^2 L_1+u_0^3 L_2
\end{equation}
{}From Eq.~(\ref{UR}) the GML function
$\beta_{GML}=(\partial u_R /\partial\ln\Lambda)_{u_0}$ is found to be
\begin{equation}
\label{GMLF}
\beta_{GML}(u_R) =-u_R^2 L_1^{\prime}+u_R^3 (L_2-L_1^2)^{\prime}
\end{equation}
where the primes in the r.h.s.~of Eq.~(\ref{GMLF}) stand for the partial
derivative w.r.t. $\ln\Lambda$. We can use symbolic form (\ref{GMLF}) for
the calculation of the GML functions for renormalized functions
$\hat\Phi(\theta,\Lambda)$,  ${\hat\Gamma}(\theta,\Lambda)$, and
${\hat V}(\theta,\Lambda)$.

The phase space analysis of the two-loop diagrams shows that the only relevant
contribution to $L_2$ in the Landau channel comes from the diagram $ZS2$. In
the  unphysical limit this contribution is zero, while in the physical limit
it is  exactly canceled by the term $-L_1^2$ which in this case is the
contribution  from the one-loop ZS diagram. Therefore,
Eqs.~(\ref{RGeq}a,~b) are true at  two-loop level as well. In the BCS
channel, the only relevant contribution to  $L_2$ in the equation for the
GML function of ${\hat V}(\theta,\Lambda)$ comes  from the diagram BCS2,
and in the same way it is canceled by the contribution $L_1$ from the BCS
diagram, leaving Eq.~(\ref{RGV}) unchanged.

It is worth noticing that the above proof of the  cancellation of higher-order
terms of the GML functions do not work in one dimension, because the phase
space restrictions in 1D are not strong enough to  remove the contributions
from all higher-loop diagrams. At the two-loop  level, the term $L_2$ in
Eq.~(\ref{GMLF}) includes the contributions from all  1PI vertex diagrams
along with the self-energy `sunrise' diagram contribution,  responsible for
the wavefunction renormalization. Thus, in 1D there is no cancellation  in
the GML functions due to renormalization.

%==============================================================================
\section{The RG equations in three dimensions}
In order to obtain the marginal coupling functions for a spherical Fermi
surface  in 3D, we follow the same procedure as in Sec.III. There are two
marginal  coupling functions:

{\it 1.~The function $\hat\Phi$.~} It is defined by
Eq.~(\ref{Fdef}). It couples a
pair of fermions with momenta $({\bf K}^1_F,{\bf K}^2_F)$ to another pair
with momenta $({\bf K}^3_F,{\bf K}^4_F)$, lying on the cone swept by a
rotation of
$({\bf K}^1_F,{\bf K}^2_F)$ around the vector ${\bf K}^1_F+{\bf K}^2_F$. Let
$\theta_{12}$ be the angle between two vectors ${\bf K}^1_F$ and
${\bf K}^2_F$, and $\phi_{13}$ the angle between the two planes defined by
the pairs $({\bf K}^1_F,{\bf K}^2_F)$ and $({\bf K}^3_F,{\bf K}^4_F)$. The
function $\hat\Phi$ may be written as
\begin{equation}
\label{Fdef3}
  \hat\Phi=\Phi^A(\theta_{12},\phi_{13})
  I^{\alpha\beta}_{\gamma\delta} +
\Phi^S(\theta_{12},\phi_{13}) T^{\alpha\beta}_{\gamma\delta}
\end{equation}
wherein the functions $\Phi^{A,S}$ have the following symmetry
properties:
\begin{mathletters}
\label{FS3}
\begin{eqnarray}
\Phi^A(\theta_{12},\phi_{13}\pm\pi) &=&
- \Phi^A(\theta_{12},\phi_{13})\\
\Phi^S(\theta_{12},\phi_{13}\pm\pi) &=&
 ~~\Phi^S(\theta_{12},\phi_{13})
\end{eqnarray}
\end{mathletters}
The minimal domain of definition allowing to recover all values of these
functions is $\theta,\phi\in [0,\pi)$.

{\it 2.~The function $\hat V$.~} It is defined by Eq.~(\ref{Vas}a), with the
symmetry properties (\ref{VSym}).

The phase space analysis shows that the RG flow is generated only in the
forward scattering amplitude channel, i.e., in the `physical' limit of the
vertex  ${\hat\Gamma}(\theta)$. All other coupling functions are RG
invariant and never mix up with the $\phi=0$ channel. Like in the 2D case, the
relevant contributions to the GML functions of ${\hat\Gamma}(\theta,t)$ and
$\hat V(\theta,t)$ come respectively from the diagrams ZS and BCS.

The spherical symmetry of the Fermi surface makes convenient an expansion of
these functions into Legendre polynomials. Using again the short-hand
notation {\bf X} for the set of coupling functions (cf Eq.~(\ref{Fourier})),
this expansion is
\begin{mathletters}
\label{Legendre}
\begin{eqnarray}
{\bf X}(\theta) &=&\sum_{l=0}^{\infty}~(2l+1)~
 {\bf X}_l~ P_l(\cos\theta)\\
{\bf X}_l &=&\frac12\int_{-1}^1 d (\cos\theta)~
{\bf X} (\theta)~ P_l(\cos\theta)
\end{eqnarray}
\end{mathletters}
Here, in the set ${\bf X}(\theta)$, $\Phi^{A,S}(\theta)$ stands for
$\Phi^{A,S}(\theta,\phi=0)$. All the calculations are conducted in the same way
as in the previous section.

The symmetry properties and the RG equations of
$\hat V$ are exactly the same as in two dimensions. The only technical
modification in the Landau channel is the sum rule for partial amplitudes
(cf. Eq.~(\ref{SUMR}))
\begin{equation}
\label{Pauli3}
\sum_{l=0}^{\infty} (2l+1)
\left[ (N-1)A_l^{\ast}+ B_l^{\ast}\right] =0
\end{equation}
The rest of the formulas and all the conclusions made in the
2D case may be repeated word for word in the 3D case.

%==============================================================================
\section{Response functions}
In this section we apply our RG technique to the study of response
functions. (For the definitions of these functions and their relationship
with physical observables, see, for instance,
Ref.~\onlinecite{Negele88}.) All the results presented below are valid
for spatial dimensions $d=2, 3$. In order to calculate the compressibility, we
consider  the density response function $\kappa({\bf Q})$, defined as follows:
\begin{equation}
\label{DR}
\delta^{(d+1)}(0)\kappa({\bf Q}) \equiv \langle \tilde{\rho}({\bf Q})
\tilde{\rho}^{\dag}({\bf Q})\rangle
\end{equation}
wherein the density operator $\rho({\bf Q})$ is
\begin{equation}
\label{DOP}
\rho({\bf Q})=\int_{({\bf 1})}\bar\psi_{\alpha}({\bf 1})
\psi_{\alpha}({\bf1}+{\bf Q})
\end{equation}
and $\tilde{\rho}({\bf Q})=\rho({\bf Q})-\langle \rho({\bf Q}) \rangle $
stands for the density fluctuation. Here ${\bf Q} \equiv ({\bf
q},\Omega_n)$, and $\Omega_n$ is the Matsubara frequency. The zeroth
component of this function $\kappa({\bf Q})$ taken in the physical limit
gives us the derivative of the particle concentration $n$ w.r.t. the
chemical potential, i.e.
\begin{equation}
\label{DNM}
\kappa \equiv
\lim_{ {\bf q}\to0}
\left[ \kappa({\bf Q})
\bigg\vert_{\Omega_n=0} \right] ={\partial n \over\partial\mu}
\end{equation}

The RG flow is simplest when obtained in the field-theory renormalization
scheme,
like in the Sec.~IIID. Notice, that in this scheme the RG parameter $\tau$ (cf.
Eq.~(\ref{tau})) runs from $\tau=0$ to $\tau=\tau_0\approx1$ (the fixed point),
and the r.h.s. of Eqs.~(\ref{RGChFl}) changes its sign. The first two terms of
the perturbative expansion for $\kappa$ give us (cf. Eq.~(\ref{ABV}))
\begin{equation}
\label{P1}
\kappa \simeq \kappa_0 -\kappa_0 \tau A_0^{(0)}
\end{equation}
wherein $\kappa_0={N \over 2}\nu_F\tau$ is the contribution from the free
part of the effective action. According to the results of the Sec.~IIID, to
obtain the equation for the RG flow of $\kappa(\tau)$, exact in this model,
it is sufficient to take into account one-loop renormalization of the
physical vertex. Introducing the auxiliary function $\bar{\kappa}(\tau)
\equiv \kappa/\kappa_0$, and using Eqs.~(\ref{RGChFl}), we get the RG
equation:
\begin{equation}
\label{RGKap}
{\partial \ln\bar{\kappa}(\tau) \over\partial\tau}=-A_0(\tau)
\end{equation}
which yields $\partial n/\partial\mu$ as its fixed point:
\begin{equation}
\label{KF}
\kappa^{\ast}={N \over 2}{\nu_F \over1+F_0}
\end{equation}
{}From the thermodynamic formula for the compressibility $K \equiv -(1/V)
(\partial V/\partial P)=(1/n^2)(\partial n/\partial\mu)$
(see, for instance, Ref.~\onlinecite{BaymP}), we easily recover the
result for the electron Fermi liquid:
\begin{equation}
\label{K0}
K={1 \over n^2}{\nu_F \over1+F_0}
\end{equation}

To find {\it the spin susceptibility} in our $SU(N)$ formalism, we consider the
flavor response function $\chi({\bf Q})$, defined as follows
\begin{equation}
\label{FR}
\delta^{(d+1)}(0)\chi({\bf Q}) \equiv {1 \over N^2-1}\langle
S_a({\bf Q}) S_a^{\dag}({\bf Q})\rangle
\end{equation}
wherein the flavor density operator
\begin{equation}
\label{FOP}
S_a({\bf Q})= {g \over 2} \int_{({\bf 1})} \lambda^a_{\alpha \beta}
\bar\psi_{\alpha}({\bf 1}) \psi_{\beta}({\bf1}+{\bf Q})
\end{equation}
and $g$ is the gyromagnetic ratio. The uniform susceptibility $\chi$ is
given by the physical limit $\chi({\bf Q}\to 0)$ (cf. Eq.~(\ref{DNM})).
Defining $\chi({\bf Q})$ in the fashion (\ref{FR}), we used the fact that in
the paramagnetic state the response is the same along all the $(N^2-1)$
directions $a$. The rest of the calculations is carried out in the same way
as above. For the auxiliary function $\bar{\chi}(\tau)
\equiv \chi /\chi_0$ ($\chi_0=\frac14 g^2 \nu_F\tau$ is the contribution in
the absence of interaction) we obtain the RG equation:
\begin{equation}
\label{RGHi}
{\partial \ln\bar{\chi}(\tau) \over\partial\tau}=-B_0(\tau)
\end{equation}
Again, we recover the FL theory result:\cite{BaymP}
\begin{equation}
\label{HiF}
\chi^{\ast}=\frac14 g^2 {\nu_F \over1+G_0}
\end{equation}
Notice, that the stability conditions for the fixed points (\ref{KF}) and
(\ref{HiF}) are just a special case ($l=0$) of the general conditions
(\ref{Stab2}).

%xxxxxxxxxxxxxxxxxxxxxxxxxxxxxxxxxxxxxxxxxxxxxxxxxxxxxxxxxxxxxxxxxxxxxxxxxx
\begin{figure}[tpb]
\input epsf.tex
\vglue 0.4cm\epsfxsize 8cm\centerline{\epsfbox{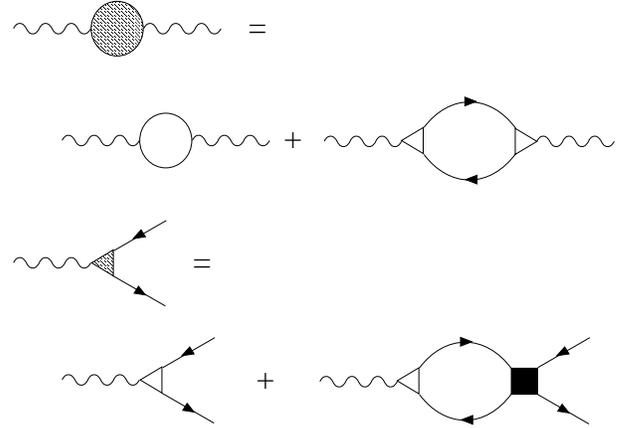}}\vglue 0.4cm
\caption{
Diagrammatic form of the recursion relations for the susceptibility $\chi(t)$
(circle) and the vertex $z(t)$ (triangle). The hatched symbols stand for the
functions at $t+dt$, whereas light ones stand for functions at $t$. The black
square stands for the vertex $\hat\Gamma(t)$. Integration over loop momenta is
carried out in an infinitesimal shell $d\Lambda(t)$.}
\end{figure}
%xxxxxxxxxxxxxxxxxxxxxxxxxxxxxxxxxxxxxxxxxxxxxxxxxxxxxxxxxxxxxxxxxxxxxxxxxx
It is interesting to show how the results (\ref{KF}) and (\ref{HiF}) can be
obtained in the Wilson-Kadanoff scheme.\cite{Note2}
Below we specialize to the electron ($N=2$) spin susceptibility, but the
compressibility may be computed in the same fashion. We add to the effective
action a source field $h({\bf Q})$, conjugated to the $z$-component of the spin
density (\ref{FOP}). For details on this approach, see, for instance,
Ref.~\onlinecite{Bourbonnais91}. The successive integration over momentum modes
generates a vertex correction $z(t)$ to the source term along with higher order
terms in the external field $h$. The effective action, at some intermediate
value of $t$ (the flow parameter), takes the form
\begin{eqnarray}
\label{Sh}
S[\psi,h,t]   &=& S[\psi,0,t] +
\int_{({\bf Q})} z(t,{\bf Q})\left[ h({\bf Q})S_3({\bf Q},t) + h.c.\right]
\nonumber\\ && ~+~
\int_{({\bf Q})} \chi(t,{\bf Q})h^*({\bf Q})h({-\bf Q}) ~+~ {\cal O}(h^3)
\end{eqnarray}
In linear response theory it is sufficient to keep track of terms up to second
order in $h$. The recursion relations for the vertex $z(t)$ and the
susceptibility $\chi(t)$ are illustrated on Fig.~2. In the physical limit ${\bf
Q}\to0$, we obtain the following pair of RG equations:
\begin{equation}
{\partial\chi\over\partial\tau} = -\frac14 g^2\nu_F z^2(\tau) \qquad\qquad
{\partial z\over\partial\tau} = z(\tau)B_0(\tau)
\end{equation}
with the initial conditions $z(\tau_0)=1$ and $\chi(\tau_0)=0$.
Using the solution for $B_0(\tau)$ which follows from (\ref{RGChFl}b),
we can easily solve this system. The uniform susceptibility, which is the
fixed point of these equations, is again given by Eq.~(\ref{HiF}).
Notice that one cannot simply compute the susceptibility (or any other response
function) from the usual diagrammatic expression at the fixed point action: one
would obtain zero at any non-zero temperature.

In order to obtain the heat capacity in this model, one does not need to
perform any complicated calculation. As explained in Sec.~IIID, tracing over
fermion modes in the vicinity of the Fermi surface in $d=2$ and 3 does not
modify the one-particle Green's function (\ref{G0}). Therefore, the only
relevant part of the free energy expressed as a functional of the total
Green's function\cite{Lutt60} is the same as for the effective action
(\ref{S0}) without the interacting part. It simply gives the Fermi liquid
result $C=\frac16\pi^2N\nu_FT$.

%==============================================================================
\section{Conclusion}

The RG approach to interacting fermions has been extended to  the case of
non-zero temperature and spin. We studied a model with  $SU(N)$-invariant
short-range effective interaction and rotationally invariant  Fermi surface.
In general, the RG approach turns out to be equivalent to Landau's mean-field
treatment of the Fermi liquid. The first set of conditions (\ref{VStab}) for
the
bare couplings of the low-energy action is nothing but Landau's theorem for the
stability of Fermi liquids against Cooper pairing at arbitrary angular
momentum.\cite{Lifshitz80} Likewise, conditions (\ref{Stab2})  for the system
to
be stable up to zero temperature are Pomeranchuk's conditions  for the
components
of the Landau interaction function. Only if both conditions (\ref{VStab}) and
(\ref{Stab2}) are fulfilled does the system reach zero temperature in the Fermi
liquid regime.
One advantage of the finite-temperature formalism is to reveal the possible
breakdown of the FL picture below some critical temperature $T_c$, whereas it
remains applicable above $T_c$.

The distinction between {\it physical} and {\it unphysical} limits as momentum
and frequency go to zero appears naturally in this finite temperature
formalism,
and differentiates the Landau function from the scattering vertex. The former
is
RG invariant, whereas the latter flows towards a fixed point  (\ref{Fix1}),
which
reproduces the well-known relationship between components of the Landau
function
and the forward scattering amplitudes. Working in the finite-temperature
formalism made explicit the equivalence of the results obtained in the
Wilson-Kadanoff and field-theory schemes of renormalization. Notice that, in
the
Wilson-Kadanoff scheme at zero temperature,  the RG flow of the physical
scattering vertex might be easily overlooked,\cite{Shankar} since special care
must be taken when the shrinking cut-off crosses the scale given by non-zero
transferred momentum.

In general, temperature plays the role of a `natural regulator' in this theory.
Firstly, it eliminates the need for an infrared cut-off in the BCS channel.
Secondly, and more important, it makes the Gell-Mann--Low functions smooth in
the Landau channel.\cite{Note3} Thus, the fixed points are smooth functions in
the $T\to0$ limit.

Response functions, like the spin susceptibility or the compressibility, are
naturally defined in the {\it physical} limit, and thus their RG flow involves
the physical vertex $\hat\Gamma(t)$, which then plays the role of a `running'
effective interaction. This makes superfluous the use of any additional
perturbative calculation\cite{Shankar} once the fixed point is reached.

%------------------------------------------------------------------------------
\acknowledgements
Stimulating conversations with C.~Bourbonnais, N.~Dupuis, A.~Ruckenstein,
A.-M.~Tremblay and Y.~Vilk are gratefully acknowledged. In particular we thank
A.-M.~Tremblay for careful reading of the manuscript. This work is supported
by NSERC and by F.C.A.R. (le Fonds pour la Formation de Chercheurs et
l'Aide\`a la Recherche du Gouvernement du Qu\'ebec).
%
%==============================================================================
% REFERENCES
%

%==============================================================================
\end{document}